\documentclass[aps,twocolumn,superscriptaddress,showpacs]{revtex4}
\usepackage{epsf,latexsym}
\usepackage{times}
\usepackage{amsfonts,epsfig}

\newcommand{\bra}[1]{\langle #1 |}
\newcommand{\ket}[1]{| #1 \rangle}

\newcommand{\qed}{$\hfill \Box$}

\newcommand{\ra}{{\rightarrow}}
\newcommand{\be}{\begin{equation}}
\newcommand{\ee}{\end{equation}}
\newcommand{\ba}{\begin{eqnarray}}
\newcommand{\ea}{\end{eqnarray}}

\newcommand{\ignore}[1]{}

\def\CC{{\rm\kern.24em \vrule width.04em height1.46ex depth-.07ex
    \kern-.30em C}}
\def\P{{\rm I\kern-.25em P}}

\def\RR{{\rm
         \vrule width.04em height1.58ex depth-.0ex
         \kern-.04em R}}

\def\bbbc{{\mathchoice {\setbox0=\hbox{$\displaystyle\rm C$}\hbox{\hbox
to0pt{\kern0.4\wd0\vrule height0.9\ht0\hss}\box0}}
{\setbox0=\hbox{$\textstyle\rm C$}\hbox{\hbox
to0pt{\kern0.4\wd0\vrule height0.9\ht0\hss}\box0}}
{\setbox0=\hbox{$\scriptstyle\rm C$}\hbox{\hbox
to0pt{\kern0.4\wd0\vrule height0.9\ht0\hss}\box0}}
{\setbox0=\hbox{$\scriptscriptstyle\rm C$}\hbox{\hbox
to0pt{\kern0.4\wd0\vrule height0.9\ht0\hss}\box0}}}}

\def\bbbz{{\mathchoice {\hbox{$\sf\textstyle Z\kern-0.4em Z$}}
{\hbox{$\sf\textstyle Z\kern-0.4em Z$}}
{\hbox{$\sf\scriptstyle Z\kern-0.3em Z$}}
{\hbox{$\sf\scriptscriptstyle Z\kern-0.2em Z$}}}}

\begin{document}
\title{Berry Phases and Quantum Phase Transitions}
\author{Alioscia Hamma}

\affiliation{ Institute for Scientific Interchange (ISI), Villa
Gualino, Viale Settimio Severo 65, I-10133 Torino, Italy}
\affiliation{Department of Chemistry,
University of Southern California, Los Angeles, CA 90089
}
\begin{abstract}
We study the connection between Berry phases and quantum phase transitions of 
generic quantum many-body systems. Consider sequences of Berry phases associated 
to sequences of loops in the parameter space whose limit is a point.
If the sequence of Berry phases does not converge to zero, then the limit point 
is a quantum critical point.  Quantum critical points are associated to failures 
of adiabaticity. We discuss the remarkable example of the anisotropic $XY$ spin 
chain in a
transverse magnetic field and detect the $XX$ region of criticality. 
\end{abstract}
\maketitle

{\em Introduction}.--- When the small variations of an external parameter cause 
a fundamental change in the macroscopic
features of a physical system at zero temperature, we say that the system has 
undergone a {\em quantum phase transition}
(QPT) \cite{sachdev}. At $T=0$, of course, phase transitions cannot be driven by 
thermal fluctuations.
Nevertheless, they can be possible because of quantum fluctuations due to the 
Heisenberg uncertainty principle.
While classical phase transitions are marked by singularities in the free-energy 
density, points of quantum phase
transitions are marked by points of non-analyticity in the energy density of the 
ground state. First-order QPT are
characterized by a discontinuity in the first derivative of the ground state 
energy, while second-order QPT have
discontinuities in the second derivative of the energy density. In these points, 
the energy spectrum is gapless and
some characteristic length scale diverges, like the equal-time correlation 
functions in the ground state. Higher order
quantum phase transitions correspond to discontinuities in the higher order 
derivatives of the energy but are of little
physical interest. In general, any continuous QPT consists in the spectrum 
becoming gapless at the
critical point \cite{sachdev}.
Another way to look at quantum phase transitions is to look at the many-body 
ground state of the system.
A quantum phase transitions results in a non analyticity of the many-body 
eigenstate. First order quantum phase
transitions arise from a level crossing in the many-body ground state, and can 
hence also happen in finite size
systems \cite{sachdev}, and in this case it is the many-body ground state to be 
discontinuous. On the other hand,
second order (or continuous) quantum phase transitions come from a higher order 
singularity in the ground state.
In order to obtain an exact transition, it is necessary the thermodynamic limit. 
The system becomes gapless in the
limit of the critical point, and excitations with arbitrary low energy are 
possible. Recently, it has been showed that the overlap between two ground 
states for different parameters is singular at the quantum critical point 
\cite{newzan}.

Quantum phase transitions are a very rich field of physics that has seen 
recently a great development from both the
theoretical and experimental point of view in systems like high-$T_c$ 
superconductors,
quantum Hall systems, fractional quantum Hall liquids, quantum magnets and so on 
\cite{sachdev}. Moreover, quantum phase
transitions can occur between phases that are not characterized by any local 
order parameter or symmetry breaking
description, but that have different {\em quantum order}, which is one of the 
most exciting and novel streams of
research in theoretical condensed matter \cite{wen}.
Recently, quantum phase transitions have been studied in the context of the 
theory of entanglement where
points of quantum phase transitions are believed to be connected to points of 
extremality for the entanglement
or its derivatives \cite{entanglement}. 

In this paper, we study the connection between Berry phases and quantum phase 
transitions.
Since the seminal paper of M. Berry \cite{berry} and the previous work by Stone 
\cite{Stone} it was clear
that Berry phases have a particular behaviour near points in which a state 
becomes degenerate with some other
state. For instance, consider the case of a quantum system described by a family 
of two-level real Hamiltonians.
The parameter space is 2-dimensional and points of degeneracy have co-dimension 
two, that is they are isolated
points. If an eigenvector acquires the phase factor $-1$ when taken around a 
circuit $C$, then the circuit $C$
encloses one degeneracy, even for arbitrarily small circuits. In many-body 
quantum systems points of degeneracy of
the Hamiltonians are quantum critical points of the first order which suggests 
the possibility that Berry phases
have some non trivial behaviour near all quantum critical points. Of course, in 
general quantum critical points
are not isolated points and this adds some complication.

The main result of this paper is that if we find a sequence of loops
in the parameter space converging to a point $\lambda_0$ and the corresponding 
Berry phases taken by an
eigenvector drawn around those loops do not converge to zero, then $\lambda_0$ 
is a quantum
critical point of some order. We call such a sequence of Berry phases {\em 
non-contractible}. Carollo and Pachos \cite{carollo}, have  first showed a 
connection between
the difference of Berry phases between first excited and ground state, and 
criticalities in the $XY$ spin chain.
In this paper, we provide a general theory of the connection between Berry 
phases and quantum phases transitions,
and, in discussing the $XY$ spin chain, we compute the Berry phase of the ground state
to prove that a non contractible sequence of Berry phases does imply a quantum 
phase transition.

{\em Topological considerations to detect degeneracies}.---
 Detecting degeneracies of a continuous Hamiltonian
that depends on a set of external parameters has been a subject
of study within a field of molecular physics for several decades
now. The first paper that addressed this problem goes back to
year $1963$ when Herzberg and Longuet-Higgins \cite{Herzberg}
discovered the sign reversal of a real electronic eigenfunctions
when continuously transported around a degeneracy. This property
was subsequently used by Longuet-Higgins
\cite{Longuet-Higgins} to state a 
topological criterion for detecting degeneracies of real
Hamiltonians. As a generalization, Stone \cite{Stone} presented
a general criterion for a genuinely complex Hamiltonian
(Hamiltonian that does not have time-reversal symmetry). A further
generalization, formulated by Johansson and Sj\"oqvist
\cite{Sjoqvist} for the case of real Hamiltonians, was shown to
be optimal.

The proof of Stone is very instructive: consider a family of Hamiltonians 
parametrized by a manifold $\mathcal
Q$ and a $3d$ volume $V\in\mathcal Q$ bounded by a closed surface $G$. We assume 
$G$ being a surface on which no
level-crossing occurs and the adiabatic theorem can be safely applied. The 
points of degeneracy in $\mathcal{Q}$
have dimension $q-3$\cite{arnold}. In the case of real Hamiltonians points of 
degeneracy have instead dimension
$q-2$ (see for example \cite{vonNeumann}). Consider now a set of loops $\{ 
\Gamma_j \}_{j=1}^M$ on $G$ that for
$M\rightarrow \infty$ continuously span the surface $G$. Then, to each loop and 
each non-degenerate eigenvalue
$i$, we can assign the corresponding Berry phase $\gamma^i_j$. We assume 
$\Gamma_1$ and $\Gamma_M$ are
infinitely small, and therefore $\gamma^i_1=0$ and $\gamma^i_M=2\pi l_i, l_i\in 
\mathbb{Z}$. If $l_i=0$, we call
the surface $G$ the \textit{phase-preserving} surface, for the value $i$. 
Otherwise, $G$ is {\em
phase-rotating}. In a phase rotating surfaces the phase factors trace a circle 
in their Argand plane. Note here
that being phase rotating or preserving is a property of the whole surface $G$ 
(and the eigenvalue $i$), and not
of a particular loop $\Gamma_j$. If we bisect the volume $V$, at least one of 
the two surfaces enclosing the two
volumes must be phase-rotating, and we can hence construct by successive 
bisection a sequence of arbitrarily
small phase rotating surfaces converging to a point $\lambda_0\in\mathcal Q$. 
Being arbitrarily small and phase
rotating this surface contains arbitrarily small loops whose associated phase 
factor are finite. Then the
wave-function must be discontinuous in $\lambda_0$ and therefore it is a point 
of degeneracy.

Can we use the Stone's test to detect quantum critical points? The answer is no 
for two reasons.
The first one is that in general quantum critical points are not isolated and 
then we are not guaranteed we can
construct a suitable surface $G$. The second reason is that only first-order 
quantum critical points can be detected
even when they are isolated. Nevertheless, if we are able to find a sequence of 
loops converging to a point such that
the geometric phase remains finite we could still argue that there is a 
level-crossing involved. Moreover, in the
thermodynamical limit, we can demonstrate that is either a level crossing or an 
avoided level crossing, as we are
going to show in the next section.

{\em Berry phases and quantum phase transitions}.--- The basic idea of this 
paper is that if Berry phases do not go to zero as we shrink the loops around a 
point $\lambda_0$ in the parameter space $\mathcal Q$, then that is a quantum 
critical point. 
Consider a simply connected $k-\dim$ manifold $\mathcal Q$ of physical 
parameters $\lambda\equiv (\lambda_1,..,\lambda_k)\in\mathcal Q$ and let $ 
H(\lambda)$ be a family of Hamiltonians on the Hilbert space $\mathcal{H}$, with 
a $\lambda-$independent domain $D$, bounded from below, and the functions 
$H(\lambda)$ with values in the Banach space of linear operators 
$L(D,\mathcal{H})$ are $k$ times continuously differentiable. The Hilbert space 
is thought to be with finite degrees of freedom. Moreover the spectra of the 
Hamiltonians are discrete. We remind that two Hamiltonians $H_0$ and $H_1$ are 
adiabatically connectible if they belong to the same connected component of 
iso-degenerate Hamiltonians \cite{hz}. This happens when the vectors of the 
degeneracies are the same. Writing the Hamiltonians in their spectral 
resolution, $H_{\alpha} = \sum_{i=1}^R \epsilon^i_{\alpha}\Pi^i_{\alpha} \, 
(\alpha = 0,1)$ ($R$ can be infinite), we can define the degeneracies vector 
associated to $H_{\alpha}$ as $D_{\alpha} := (\mbox{tr} 
\Pi^1_{\alpha},...,\mbox{tr} \Pi^R_{\alpha})$, ordering the components according 
to the crescent order of the corresponding eigenvalues. The fact that two 
Hamiltonians $H_0$ and $H_1$ are adiabatically connectible does not mean that we 
can adiabatically turn one into the other just changing the parameters in 
$\mathcal Q$. For this we need an {\em adiabatic surface}: a surface $S$ in the 
parameters space $\mathcal Q$ is said {\em adiabatic} if i) $D(\lambda)$ is 
constant on $S$, and ii) The projectors $\Pi^i(\lambda)$ are of class $C^3$. On 
an adiabatic surface, for every curve there is a parametrization such that the 
adiabatic theorem holds. 
%%%%%%%%%%%%%%%%%%%%%%%%%%%%%%%%%%%%%%%%%%%%%%%%%

When the adiabatic curvature diverges in a point, no adiabatic evolution is 
possible for some trajectories through that point: let $S-\{\lambda_0\}$ be an 
adiabatic surface in $\mathcal Q$. Consider the adiabatic curvature \cite{berry} 
defined by $F^j = F^j_{\mu\nu}d\mu\wedge d\nu = -\mbox{Im} 
(\partial/\partial\lambda^{\mu}\bra{j(\lambda)}) \cdot 
(\partial/\partial\lambda^{\nu}\ket{j(\lambda)}) d\lambda^{\mu}\wedge 
d\lambda^{\nu}$. Where $\ket{j(\lambda)}$ is the $j-$th eigenvector of 
$H(\lambda)$. If the curvature diverges in $\lambda_0$, i.e., 
$\lim_{\lambda\ra\lambda_0 }\Vert F^j \Vert = +\infty$, then the surface is non 
adiabatic in $\lambda_0$. Indeed, consider the projector $\Pi^j(\lambda) = 
\ket{j(\lambda)}\bra{j(\lambda)}$ and its derivatives $d/d\lambda^{\mu} \Pi^j 
(\lambda)= d/d\lambda ^{\mu} (\ket{j(\lambda)}\bra{j(\lambda)})$. The divergence 
of $F$ implies that for some $\mu$,
\be
\lim_{\lambda\rightarrow\lambda_0}\Vert\frac{d}{d\lambda^{\mu} 
}\Pi^j(\lambda)\Vert = +\infty
\ee
Now consider a regular curve $\lambda_s: s\in[0,1]\rightarrow \lambda 
(s)\in\mathcal Q$ such that $\lambda (s_0)=\lambda_0$ with $s_0\in [0,1]$. For 
every parametrization of this curve, we obtain $\lim_{s\rightarrow s_0}
\Vert \frac{d\Pi^n(s)}{ds}\Vert = \lim_{s\rightarrow s_0}\Vert 
\frac{d\Pi^n(\lambda)}{d\lambda^{\mu} }\Vert \,|\frac{d\lambda^{\mu} (s)}{ds}| = 
+\infty
$, which means that $\Vert d\Pi^n/ds\Vert$ diverges along every regular curve in 
the limit $s\rightarrow s_0$ and hence the Hamiltonians on the curve $\lambda_s$ 
are not adiabatically connectible in $\lambda_0$.

%%%%%%%%%%%%%%%%%%%%%%%%%%%%%%%%%%%%%%%%%%%%%%%%%%
{\em Proposition 1}.--- Consider a $k-$differentiable family of Hamiltornians 
$H(\lambda)$ such that $S\subset\mathcal Q$ is an adiabatic surface everywhere 
but in the point $\lambda_0$. We also assume that, for the ground state $\ket{j}$,  $\Vert \partial_{\mu}H\ket{j}\Vert $ is finite for every $\mu$. Let
us consider  a sequence of loops $\Gamma_r$ converging to some point 
$\lambda_0\in\mathcal Q$, that is,
$\lim_{r\rightarrow\infty}\Gamma_r = \lambda_0$  and such that along every 
$\Gamma_r$ we can apply the adiabatic theorem. Let us compute, for the ground 
state $\ket{j}$, the corresponding Berry phases $\gamma_r^j$ and let the
associated sequence of Berry phases $\{\gamma_r^j\}$ be {\em
non-contractible}, i.e., $\lim_{r\rightarrow\infty}\gamma_r^j\ne 0$.
Then $S$ is not adiabatic in $\lambda_0$ and $\lambda_0$ is a quantum critical 
point. Notice that we do not need to be
able to apply the adiabatic theorem in $\lambda_0$, which is the candidate 
critical point.\\
{\em Proof}.---  Consider the curvature associated to the Berry connection
$F^j= dA^j = F^j_{\mu\nu}d\mu\wedge d\nu$. 
The Berry phase associated to the loop $\Gamma$ for the ground state 
$\ket{j(\lambda)}$ is given by the flux of $F$ through the surface $S$ bounded 
by $\Gamma$:
$
\gamma = - \int_S F^j
$. 
If we shrink $S$ to the point $\lambda_0$, the flux goes to zero unless $F^j$ 
diverges in norm in $\lambda_0$. Therefore $S$ is not adiabatic in $\lambda_0$. 
Using the resolution of the identity, we can write $F^j_{\mu\nu}$ as
\be
F^j_{\mu\nu} = \sum_{m\ne j}\frac{\bra{j}\partial_{\mu} 
H \ket{m}\bra{m} \partial_{\nu} 
H \ket{j}}{(E_m - E_j)^2}
\ee
Now let $\Delta E = \inf_{m}\vert E_j - E_m\vert$ and $A^{\mu}_{jm} =   
\bra{j}\partial_{\mu}H\ket{m}$. We have
\ba\label{fin}
\nonumber
\vert F^j_{\mu\nu}\vert &\le& \sum_{m\ne j}\frac{\vert A^{\mu}_{jm}
A^{\nu}_{mj}\vert}{(E_m-E_j)^2}\le \frac{\sum_{m}\vert A^{\mu}_{jm}
A^{\nu}_{mj}\vert}{(\Delta E)^2}\\
&\le& \frac{1}{(\Delta E)^2}\Vert \partial_{\mu}H\ket{j}\Vert \cdot \Vert \partial_{\nu}H\ket{j}\Vert 
\ea
where the last inequality follows from the Cauchy-Schwarz inequality. The numerator of the above expression is finite and thus if $F^j_{\mu\nu}$ 
diverges in norm in $\lambda_0$ then $\Delta E$ must go to zero in the same 
point.
Since $\ket{j}$ is the ground state, then $\lambda_0$ is a quantum critical point 
because we have a level crossing and thus a first order QPT. This proposition is 
easily generalizable to Hamiltonians that have a continuous spectrum, as long as 
the low energy sector is discrete. We do not need to have infinite degrees of 
freedom for a first order QPT.

Nevertheless, continuous phase transitions are possible in the thermodynamic 
limit. We want to show that the divergence of the curvature still implies 
gaplessness for a system in the thermodynamic limit. Consider a sequence of  
families of Hamiltonians $\{ H_n(\lambda)\}_{n\in\mathbb N}$. Here $n$ means the 
number of degrees of freedom, so the thermodynamic limit is the system described 
by $\
n\ra\infty$. Let us assume that for every finite $n$, the surface $S$ is 
adiabatic. Then no first order QPT is possible and for every finite $n$,  
$\lim_{r\rightarrow\infty}\gamma_r^j(n) = 0$. But it could happen that for $n\ra 
+\infty$ some gap with the ground state closes, thus leading to a continuous 
QPT. To perform the thermodynamic limit, we need intensive quantities. So if the 
normalization of the ground state scales like $\mathcal N (n)$, then we need to 
use the intensive cuvature $\tilde{F}^j = F^j/\mathcal N(n)^2$. Now let the 
sequence be non-contractible only in the thermodynamic limit
\be
\lim_{r\rightarrow\infty}\lim_{n\ra\infty}\mathcal N(n)^{-2}\gamma_r^j(n) \ne0
\ee
 Then $\vert \tilde{F}^j_{\mu\nu} \vert$ diverges in $\lambda_0$ and because the 
numerator of Eq.(\ref{fin}) will now diverge like $\mathcal N(n)^2$, this implies 
again that $\lim_{\lambda\ra\lambda_0}\Delta E =0$.\qed

%%%%%%%%%%%%%%%%%%%%%%%%%%%%%%%%%%%%%%%%%%%%%%%%%%%
{\em Example: the XY spin chain}.---  In this section we show a
remarkable example illustrating how a non-contractible sequence of
Berry phases reveals quantum criticalities. Let us consider as an example the 
$XY$
model. It represents one-dimensional anisotropic spin chain
subjected to an external field along the $z$ axis. The system is
given by the following Hamiltonian $H(\gamma, \lambda)$: 
\be
H(\lambda, \gamma) = -\sum_{i = -M}^{M} \left(
\frac{1+\gamma}{2}\hat{\sigma}_{i}^x \hat{\sigma}_{i+1}^x +
\frac{1-\gamma}{2} \hat{\sigma}_{i}^y \hat{\sigma}_{i+1}^y +
\lambda\hat{\sigma}_{i}^z\right) 
\ee
Here, the parameter
$\gamma \in \left[0,+\infty \right[$ represents an anisotropy in the
next-neighbor spin-spin interaction, while $\lambda \in
\mathbb{R}$ is an external field. The operators $\hat{\sigma}_{i}^{\alpha},\, 
\alpha \in \{
x,y,z\}$ are the usual Pauli operators. We will consider the more general 
Hamiltonian $H(\lambda, \gamma, \phi) = \hat{g}(\phi)
H(\lambda, \gamma) \hat{g}^{\dagger}(\phi)$, where
$\hat{g}(\phi) = \prod_{i = -M}^M e^{i\hat{\sigma}_i^z \phi / 2}$,
with $\phi \in [0,2\pi)$. This way, the parameter space is the
simply connected three dimensional Euclidean space
$\mathcal{Q}\equiv\mathbb{R}^3$. This Hamiltonian can be
diagonalized by successively applying Jordan-Wigner, Fourier and
Bogoliubov transformation (for details, see
\cite{sachdev}), after which we obtain the free fermion
Hamiltonian $H(\lambda, \gamma, \phi) = \sum _{k =
-M}^M\Lambda_k\hat{b}^{\dagger}_k \hat{b}_k$. The energies
$\Lambda_k$ of one-particle excitations, defined by the operators
$\hat{b}_k$, are
$\Lambda_k=\sqrt{\varepsilon_k^2+\gamma^2\sin^2\frac{2\pi k}{N}}$,
with $\varepsilon_k=\cos\frac{2\pi k}{N}-\lambda$. The ground
state of this Hamiltonian is given as a product of qubits in the
following way: 
\be \label{ground_state} 
\ket{g} = \bigotimes_{
k=1}^M\left( \cos\frac{\theta_k}{2} \ket{0}_k\ket{0}_{-k} 
+ie^{2i\phi}\sin\frac{\theta_k}{2}\ket{1}_k\ket{1}_{-k} \right) 
\ee
where $\ket{0}_k$ and $\ket{1}_k$ are the vacuum and first excited
states of the mode $\hat{d}_k$ introduced by the Fourier
transformation, while $\hat{b}_k=\cos \frac{\theta_k}{2}\hat{d}_k
-ie^{2i\phi}\sin\frac{\theta_k}{2}\hat{d}_{-k}^{\dagger}$ and $
\cos\theta_k = \varepsilon_k/\Lambda_k =
(\cos\frac{2\pi k}{N}-\lambda)\left((\cos\frac{2\pi
k}{N}-\lambda)^2+\gamma^2\sin^2\frac{2\pi k}{N}\right)^{-1/2} $.

The $XY$ model exhibits three regions of criticality. The $XX$ region of
criticality is defined by the following conditions: $\gamma = 0$
and $\lambda \in (-1,1)$; the $XY$ regions of criticality are
given by the two planes defined by $\lambda = \pm 1$, and the
Ising region of criticality is defined by $\gamma = 1$.

We want to show that the $XX$ region of criticality is characterized
by the existence of a non-contractible Berry phase. The critical
phenomena in the $XX$ region are given by the existence of the
gapless excitations defined by $\hat{b}^{\dag}_{k_0}$, where
$\cos\frac{2\pi k_0}{N}=\lambda$, which is a consequence of the
condition $\Lambda_{k_0}=0$. Note that, as  $k_0 \in\{1, \ldots
M\}$, the above condition can, for a general $\lambda \in (-1,1)$,
satisfied in thermodynamical limit ($N \rightarrow \infty$) only. In the 
continuum limit we define
$x\equiv \lim_{N\rightarrow\infty} 2\pi k/N$ and hence we have $\cos x_0 = 
\lambda$. So, as long as  $M$ remains finite, we have
\footnote{
For some particular values of $\lambda$, of course, there is a $k$ such that 
$\cos\theta_k = 0$,
but this happens for a set of values of the parameters with null measure and we 
neglige it.
}
\be\label{limcos}
\lim_{\gamma\rightarrow 0}\cos\theta_k = \pm 1
\ee
For $\gamma = 0$, the ground state
is hence given by the set of conditions: $\cos\theta_k= \mbox{sgn}
(\varepsilon_k)=\pm 1$, i.e., $\ket{g(\lambda, \gamma
= 0, \phi; M)} = \bigotimes_{ k<k_0} \ket{0}_k\ket{0}_{-k}
\bigotimes_{ k>k_0} \ket{1}_k\ket{1}_{-k}$. 
The Berry phase acquired by the ground state of the system upon
adiabatic change of the parameter $\phi$ from $0$ to $\pi$ (with
$\lambda$ and $\gamma$ fixed) can be easily calculated if we
observe that the ground state has a formal structure of a tensor
product of pure states of $M$ qubits, each defined by the angles
$\theta_k$ and $\phi$. Then, the overall Berry phase is just the
sum of $M$ one-qubit Berry phases (note that, although $\phi \in
[0,\pi]$, each qubit state makes one complete loop in parameter
space due to the $2$ factor in the exponent):
\be
\label{Berry-ground_state}
\varphi(M)=-i\int_0^{\pi}
\bra{g} \frac{\partial}{\partial \phi}\ket{g}=
\sum_{k=1}^M\pi (1- \cos \theta_k)=\sum_{k=1}^M
\varphi_k
\ee
where by $\varphi_k\equiv \pi(1- \cos \theta_k)$ we
denote a single-qubit Berry phase, defined by the number $k$. Using the 
expression for the ground state and Eq.(\ref{limcos}) we find that, for all $k$ 
and every finite $M$,
\be\label{phik}
\lim_{\gamma\rightarrow 0}\varphi_k = 0, 2\pi
\ee
and henceforth $\varphi(\ket{g}) = 0,2\pi $. So for finite $M$, the Berry phase 
acquired by the ground
state is always trivial. Nevertheless, in the thermodynamical limit, there is 
always a
solution for $\cos\theta_{k_0} = 0 \Leftrightarrow\cos x_0
=\lambda $. This implies that $\varphi_{k_0} = \pi$ for every $\gamma>0$ and
 hence $\lim_{\gamma\rightarrow 0}\varphi_{k_0} = \pi$. Direct calculation shows 
that Eq.(\ref{Berry-ground_state}) and (\ref{phik}) imply \be
 \lim_{\gamma\rightarrow 0}\lim_{M\rightarrow\infty} \frac{1}{M}\varphi (M)\ne0
\ee
so the sequence $\{ \varphi_{\gamma_n}(M)\}_{n\in\mathbb N}$ is non contractible 
in the thermodynamic limit and therefore the
 $XX$ criticality is detected by a non-contractible Berry phase. In a recent 
work, Carollo and Pachos
 \cite{carollo} have computed the difference of the Berry phases acquired by the 
ground state and the
 first excited state for loops around the $XX$ region of criticality and found 
that this relative Berry
 phase is non trivial and converges to $-\pi$ as $\gamma$ tends to zero. This is 
because this relative
 Berry phase is, in the limit $\gamma\rightarrow 0$, the one (with opposite 
sign) taken by the single
 equatorial qubit defined by $k_0$.

{\em Conclusions and perspectives}.--- In a critical system near zero 
temperature quantum fluctuations,
due to the Heisenberg principle, can drive transitions between one phase and 
another, that is from one
internal order to another one. In this paper, we have shown that the non 
contractibility of
Berry phases for the ground state is associated to quantum phase transitions and 
to a failure of adiabaticity because of the divergence of the Berry curvature.

It is very interesting that quantum physics revealed that internal orders
of the matter are not always described by Landau-Ginzburg or local order 
parameters and breaking of
symmetry \cite{landau}, as it has been pointed out since the discovery of 
fractional quantum Hall
liquids \cite{fqhs, wenpapers}. At very low temperatures, the internal order of 
strongly correlated
systems like systems with strongly correlated electrons consists of the pattern 
of how the electrons
move with respect to each other and there can be different phases with the very 
same symmetries. Novel notions like quantum and topological order have been
proposed \cite{wenpapers}.The world of quantum
orders is a very rich and largely unexplored field of quantum physics and it 
would be intriguing to use Berry phases to distinguish different quantum or 
topological orders, like a string-condensed
state from one without string condensation.

{\em Acknowledgments}.--- After the main part of this work was completed, the 
author became aware of the work of S.-L. Zhu who had independently conjectured 
the general connection between Berry phases and QPT \cite{zhu}. The author 
thanks N. Paunkovi\'c, P. Zanardi and K. Khodjasteh for important and enlightening discussions.

\end{document}